\documentclass[final]{IEEEtran}

\usepackage[encapsulated]{CJK}
\usepackage{ucs}
\usepackage[utf8x]{inputenc}
\usepackage[cmex10]{amsmath}
\usepackage{amsmath,amssymb,amscd,bbm,amsthm,mathrsfs,dsfont}
\usepackage{algorithmic,algorithm}
\usepackage{mdwmath}
\usepackage{mdwtab}
\usepackage{bm,upgreek}
\usepackage{cite}
\usepackage{rotating,graphics,psfrag,epsfig}
\usepackage{array}
\usepackage{booktabs}
\usepackage{indentfirst}
\usepackage{subfigure}
\usepackage{lipsum,fancyhdr,lastpage,refcount}
\usepackage[hyphens]{url}
\usepackage{fixltx2e,dblfloatfix}
\usepackage{blindtext}
\usepackage[T1]{fontenc}
\usepackage{color}

\IEEEoverridecommandlockouts

\let\oldnl\nl
\newcommand{\nonl}{\renewcommand{\nl}{\let\nl\oldnl}}

\begin{document}

\title{Resource Awareness in Unmanned Aerial Vehicle-Assisted Mobile-Edge Computing Systems}

\author{\IEEEauthorblockN{Xianfu Chen\IEEEauthorrefmark{1}, Tao Chen\IEEEauthorrefmark{1}, Zhifeng Zhao\IEEEauthorrefmark{2}, Honggang Zhang\IEEEauthorrefmark{3}, Mehdi Bennis\IEEEauthorrefmark{4}, and Yusheng Ji\IEEEauthorrefmark{5}}

{\IEEEauthorrefmark{1}VTT Technical Research Centre of Finland Ltd, Finland}\\
{\IEEEauthorrefmark{2}Research Center for Intelligent Networks, Zhejiang Lab, Hangzhou, China}\\
{\IEEEauthorrefmark{3}College of Information Science and Electronic Engineering, Zhejiang University, China}\\
{\IEEEauthorrefmark{4}Centre for Wireless Communications, University of Oulu, Finland}\\
{\IEEEauthorrefmark{5}Information Systems Architecture Research Division, National Institute of Informatics, Tokyo, Japan}
\vspace{-.39cm}

\thanks{This work has been submitted to the IEEE for possible publication. Copyright may be transferred without notice, after which this version may no longer be accessible.}
}

\maketitle

\begin{abstract}

This paper investigates an unmanned aerial vehicle (UAV)-assisted mobile-edge computing (MEC) system, in which the UAV provides complementary computation resource to the terrestrial MEC system.
The UAV processes the received computation tasks from the mobile users (MUs) by creating the corresponding virtual machines.
Due to finite shared I/O resource of the UAV in the MEC system, each MU competes to schedule local as well as remote task computations across the decision epochs, aiming to maximize the expected long-term computation performance.
The non-cooperative interactions among the MUs are modeled as a stochastic game, in which the decision makings of a MU depend on the global state statistics and the task scheduling policies of all MUs are coupled.
To approximate the Nash equilibrium solutions, we propose a proactive scheme based on the long short-term memory and deep reinforcement learning (DRL) techniques.
A digital twin of the MEC system is established to train the proactive DRL scheme offline.
Using the proposed scheme, each MU makes task scheduling decisions only with its own information.
Numerical experiments show a significant performance gain from the scheme in terms of average utility per MU across the decision epochs.

\end{abstract}


\section{Introduction}
\label{intr}

Mobile-edge computing (MEC), which provides computing capabilities within the radio access networks (RANs) in close proximity to the mobile users (MUs), is a promising paradigm to address the tension between computation-intensive applications and resource-constrained mobile devices \cite{Mao17}.
By offloading computation tasks to the resource-rich MEC cloud, not only the computation qualities of service and experience can be greatly improved, but also the capability of a mobile device can be augmented for running a variety of resource-demanding applications.
Recently, there are a number of related works on designing computation offloading schemes.
For example, in \cite{Wang18}, Wang et al. proposed a Lagrangian duality method to minimize the total energy consumption in a computation latency constrained wireless powered multiuser MEC system.
In \cite{Liu17}, Liu et al. studied the power-delay tradeoff for a MEC system using the Lyapunov optimization technique.
In our priori work \cite{Chen19J}, the infinite time-horizon Markov decision process (MDP) framework was used to model the problem of computation offloading for a MU in an ultra-dense RAN and to solve the optimal policies, we proposed the deep reinforcement learning (DRL) based schemes.

Offloading the input data of a task from the mobile device of a MU to the MEC cloud requires wireless transmissions, which account for the dynamics from the surrounding environment.
Particularly, the time-varying channel qualities due to the MU mobility in turn limits the computation performance \cite{Chen19}.
Because of among others, the low deployment cost, the flexibility and the line-of-sight (LOS) connections, unmanned aerial vehicles (UAVs) are expected to play a significant role in advancing the future wireless networks \cite{Moza19}.
Leveraging the UAV technology in a MEC system has been shown to be substantial.
In \cite{Hu19}, Hu et al. put forward an alternating algorithm to minimize the weighted sum energy consumption for a UAV-assisted MEC system.
In \cite{Zhou18}, Zhou et al. investigated a UAV-enabled wireless-powered MEC system and derived alternating algorithms to solve the computation rate maximization problems under both the partial and the binary computation offloading modes.
However, most of the existing literature is basically based on a finite time-horizon.

In this paper, we concentrate on a three-dimensional UAV-assisted MEC system, in which a UAV is implemented as a complementary computing server flying in the air.
That is, in addition to local computation execution, each MU in the system can also offload a computation task to the UAV or to the MEC cloud via one of the base stations (BSs) in the RAN.
The UAV can co-execute the computation tasks of the MUs by creating isolated virtual machines (VMs) \cite{Lian19}.
Sharing the same physical UAV platform causes I/O interference, leading to computation rate reduction for each VM.
Under this context, the MUs compete to schedule local and remote task computations with the awareness of environmental dynamics.
The aim of each MU is to maximize the expected long-term computation performance.
The non-cooperative interactions among the MUs are modeled as a stochastic game.
Solving a Nash equilibrium (NE) of the stochastic game needs complete information exchange among the MUs, which is practically overwhelming.
Motivated by recent advances in recurrent and deep neural networks, we propose a proactive DRL scheme, enabling each MU to behave at an approximated NE only with local information \cite{Chen19A, Napa19}.
Furthermore, we establish a digital twin of the MEC system to get over the hurdle of training the neural networks \cite{Dong19}.
To the best of our knowledge, there does not exist a comprehensive study on stochastic resource awareness among the non-cooperative MUs in a UAV-assisted MEC system.

\section{System Descriptions and Assumptions}
\label{sysm}

\begin{figure}[t]
  \centering
  \includegraphics[width=16pc]{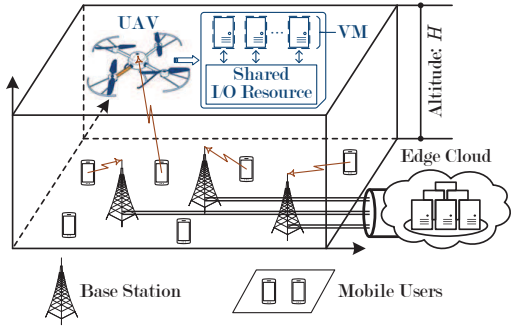}
  \caption{Illustration of an unmanned aerial vehicle (UAV)-assisted mobile-edge computing system (VM: virtual machine.).}
  \label{systMode}
\end{figure}

As illustrated in Fig. \ref{systMode}, we focus on a three-dimensional scenario, in which a terrestrial MEC system is assisted by a UAV.
The UAV hovers in the air at a fixed altitude of $H$ (in meters) \footnote{This work assumes that the power of the UAV is supplied by laser charging \cite{Liu19}. Hence the UAV is able to operate over the long run.}.
The terrestrial MEC system consists of a set $\mathcal{B} = \{1, \cdots, B\}$ of BSs, which are connected via wired links to the computing cloud at the edge.
To ease analysis, we use a common finite set $\mathcal{L}$ of locations (i.e., small two-dimensional non-overlapping areas)\footnote{Each location or small area can be characterized by uniform wireless communication conditions \cite{Chen19}.} to denote both the terrestrial service region covered by the BSs and the region of the UAV mapped vertically from the air to the ground.
In the system, a set $\mathcal{K}$ of MUs coexist and generate sporadic computation tasks over the infinite time-horizon, which is discretized into decision epochs.
Each epoch is assumed to be of equal duration $\delta$ (in seconds) and indexed by an integer $j \in \mathds{N}_+$.

\subsection{Mobility Model}

We apply the smooth-turn mobility model with a reflecting boundary to simulate the UAV trajectory \cite{Wan13}.
In this model, the UAV maintains a constant forward speed but randomly changes the centripetal acceleration.
Let $L_{(\mathrm{UAV})}^j \in \mathcal{L}$ be the mapped terrestrial location of the UAV during a decision epoch $j$.
With regards to the MUs, their movements are modelled using a boundary Gauss-Markov mobility model \cite{Xi19}.
Specifically, the location $L_{(\mathrm{MU}), k}^j \in \mathcal{L}$ of each MU $k \in \mathcal{K}$ during each decision epoch $j$ is determined by both the location $L_{(\mathrm{MU}), k}^{j - 1}$ at epoch $j - 1$ and the velocity during epoch $j$, while the velocity of a MU during a decision epoch depends on the velocity during the previous epoch only.

\subsection{Task Model}

The computation task arrivals at the MUs are assumed to be independent and identically distributed sequences of Bernoulli random variables with a common parameter $\lambda \in [0, 1]$.
More specifically, we choose $A_k^j \in \{0, 1\}$ to be the task arrival indicator for a MU $k \in \mathcal{K}$, that is, $A_k^j = 1$ if a computation task is generated at MU $k$ in the end of epoch $j$ and otherwise, $A_k^j = 0$.
Then, $\mathbb{P}(A_k^j = 1) = 1 - \mathbb{P}(A_k^j = 0) = \lambda$, $\forall k \in \mathcal{K}$, where $\mathbb{P}(\cdot)$ denotes the probability of the occurrence of an event.
We let $\mu$ (in bits) and $\vartheta$ represent, respectively, the input data size and the number of CPU cycles required to accomplish one input bit of a computation task.
The arrived but not processed tasks will be queued at the buffer of a MU.
A computation task can be either computed locally at the device of the MU or executed remotely (at the UAV or the MEC cloud).
We let $X_k^j \in \{0, 1\}$ and $F_k^j \in \mathcal{B} \cup \{0, B + 1\}$ denote the local and remote computation task scheduling decisions of MU $k$ at each decision epoch $j$.
That is, $X_k^j = 1$ if MU $k$ sends a computation task to the local CPU and otherwise, $X_k^j = 0$, while if MU $k$ offloads the computation task to the UAV, $F_k^j = B + 1$, or to the MEC cloud via one of the BSs, $F_k^j = b$ ($b \in \mathcal{B}$) and otherwise, $F_k^j = 0$.
Hence the task queue dynamics of MU $k$ can be expressed as
\begin{align}\label{taskQueu}
  Q_k^{j + 1} = \max\!\left\{Q_k^j - X_k^j - \mathds{1}_{\left\{F_k^j > 0\right\}}, 0\right\} + A_k^j,
\end{align}
where $Q_k^j$ is the number of computation tasks in the task buffer of MU $k$ at the beginning of decision epoch $j$ and $\mathds{1}_{\{\cdot\}}$ is an indicator function that equals $1$ if the condition is satisfied and $0$, otherwise.
In this work, we assume a large enough buffer capacity for a MU to avoid the buffer overflows.

\subsection{Computation Model}

The UAV complements the terrestrial MEC system with the computation resource from the air.
By strategically offloading the computation tasks to the UAV or the MEC cloud via one of the BSs for remote execution, the MUs can expect a significantly improved computation experience.

\subsubsection{Local Computation}

When a computation task is scheduled for processing locally at the mobile device of a MU $k \in \mathcal{K}$ during a decision epoch $j$, i.e., $X_k^j = 1$, the number of needed epochs can be calculated as $\Delta = \lceil (\mu \cdot \vartheta) / (\rho \cdot \delta) \rceil$, where $\lceil \cdot \rceil$ means the ceiling function and we assume that the local CPU of a MU operates at frequency $\rho$ (in Hz).
We describe the local processing state $S_{(\mathrm{MU}), k}^j \in \{0, 1, \cdots, \Delta\}$ at a decision epoch $j$ using the number $S_{(\mathrm{MU}), k}^j$ of remaining epochs to finish the computation task.
For local computation during an epoch $j$, the processing delay experienced by MU $k$ is given by
\begin{align}\label{locaProcDela}
  D_{(\mathrm{MU}), k}^j =
  \left\{\!\!\!
  \begin{array}{l@{~}l}
    0,                                                                                   & \mbox{if } S_{(\mathrm{MU}), k}^j = 0; \\
    \dfrac{\mu \cdot \vartheta - \left(\Delta - 1\right) \cdot \delta \cdot \rho}{\rho}, & \mbox{if } S_{(\mathrm{MU}), k}^j = 1; \\
    \delta,                                                                              & \mbox{if } S_{(\mathrm{MU}), k}^j > 1,
  \end{array}
  \right.
\end{align}
and the resulted energy consumed by the mobile device of MU $k$ then is
\begin{align}\label{locaEnerCons}
 & E_{(\mathrm{MU}), k}^j =                                                                                                          \nonumber\\
 & \left\{\!\!\!
    \begin{array}{l@{~}l}
     0,                                                                                  & \mbox{if } S_{(\mathrm{MU}), k}^j = 0;    \\
     \tau \cdot \left(\mu \cdot \vartheta - \left(\Delta - 1\right) \cdot \delta \cdot \rho\right)
          \cdot \left(\rho\right)^2,                                                     & \mbox{if } S_{(\mathrm{MU}), k}^j = 1;    \\
     \tau \cdot \delta \cdot \left(\rho\right)^3,                                        & \mbox{if } S_{(\mathrm{MU}), k}^j > 1,
    \end{array}
   \right.
\end{align}
where $\tau$ is the effective switched capacitance that depends on the chip architecture of a mobile device \cite{Burd96}.
%
%Consequently, the local processing state of MU $k$ is updated as $S_{(\mathrm{MU}), k}^{j + 1} = \max\{S_{(\mathrm{MU}), k}^j - 1, 0\}$.

\subsubsection{Remote Execution}

For remote computation execution, a MU has to be first associated with a BS or the UAV until the task is accomplished.
Let $I_k^j \in \mathcal{B} \cup \{B + 1\}$ be the association state of each MU $k \in \mathcal{K}$ during a decision epoch $j$, namely, $I_k^j = b \in \mathcal{B}$ if MU $k$ is associated with a BS $b$ and if MU $k$ is associated with the UAV, $I_k^j = B + 1$.
Then
\begin{align}
  I_k^j = i \cdot
  \mathds{1}_{\left\{\left\{F_k^j = i\right\} \bigvee \left\{\left\{F_k^j = 0\right\} \bigwedge \left\{I_k^{j - 1} = i\right\}\right\}\right\}},
\end{align}
where $i \in \mathcal{B} \cup \{B + 1\}$, while $\vee$ and $\wedge$ mean, respectively, logic OR and logic AND.
When $I_k^j \neq I_k^{j - 1}$, which may happen only when $F_k^j > 0$\footnote{If a MU $k \in \mathcal{K}$ does not offload a task at the beginning of a decision epoch $j$, the association state remains unchanged, i.e., $I_k^j = I_k^{j - 1}$. In this case, no handover will be triggered.}, a handover among the BSs and the UAV is hence triggered \cite{Chen19J}.
We assume that the energy consumption during the occurrence of one handover is negligible at MU $k$ but the incurred delay is $\zeta$ (in seconds).
During a decision epoch $j$, MU $k$ experiences the average channel power gains $G_{b, k}^j = g_{(\mathrm{BS})}(L_{(\mathrm{MU}), k}^j, L_{(\mathrm{BS}), b})$ for the link between MU $k$ and BS $b$ and $G_{(\mathrm{UAV}), k}^j = g_{(\mathrm{UAV})}(L_{(\mathrm{MU}), k}^j, L_{(\mathrm{UAV})}^j, H)$ for the link between MU $k$ and the UAV, which are determined by the physical distances.

At the beginning of a decision epoch $j$, if a MU $k \in \mathcal{K}$ lets the MEC cloud execute a computation task, all input data needs to be offloaded via a BS $F_k^j = b \in \mathcal{B}$, for which the achievable data rate can be written as $R_{b, k}^j = W \cdot \log_2(1 + (G_{b, k}^j \cdot P_k) / (W \cdot \sigma^2))$, where $W$ is the frequency bandwidth exclusively allocated to a MU, $P_k$ is the transmit power and $\sigma^2$ is the noise power spectral density.
We use $T_{(\mathrm{BS}), k}^j \in [0, \mu]$ to denote the local transmission state of MU $k$ at the beginning of a decision epoch $j$, which indicates the remaining amount of input data to be transmitted for the task.
Hence the transmission delay\footnote{The transmission delay includes the delay during the handover procedure.} and the energy consumption during epoch $j$ are calculated as $Y_{(\mathrm{BS}), k}^j = \min\{T_{(\mathrm{BS}), k}^j / R_{b, k}^j + \zeta \cdot \mathds{1}_{\{I_k^j \neq I_k^{j- 1}\}}, \delta\}$ and $E_{(\mathrm{BS}), k}^j = P_k \cdot (Y_{b, k}^j - \zeta \cdot \mathds{1}_{\{I_k^j \neq I_k^{j- 1}\}})$.
In this paper, we assume that the BSs are connected using the wired links to the MEC cloud, which is of rich computation resource.
We ignore the round-trip delay between the BSs and the MEC cloud as well as the time consumed for processing a computation task at the MEC cloud.
Further, the time consumed by the selected BS (or the UAV in the following) to send back the computation result is negligible due to the fact that the size is much smaller than the input data of a computation task \cite{Chen16}.

Similarly, if a MU $k \in \mathcal{K}$ offloads a computation task to the UAV for processing at a decision epoch $j$, namely, $F_k^j = B + 1$,
the time\footnote{After receiving all the input data of a computation task during a current decision epoch, the UAV starts to process from the subsequent decision epoch since the VMs are created at the beginning of an epoch \cite{Lian19}.}
and the energy consumed during each decision epoch $j$ turn to be
$Y_{(\mathrm{UAV}), k}^j = \delta \cdot \mathds{1}_{\{T_{(\mathrm{UAV}), k}^j > 0\}}$
and
$E_{(\mathrm{UAV}), k}^j = P_k \cdot (\min\{T_{(\mathrm{UAV}), k}^j / R_{(\mathrm{UAV}), k}^j + \zeta \cdot \mathds{1}_{\{I_k^j \neq I_k^{j- 1}\}}, \delta\} - \zeta \cdot \mathds{1}_{\{I_k^j \neq I_k^{j- 1}\}})$, respectively,
where $R_{(\mathrm{UAV}), k}^j = W \cdot \log_2(1 + (G_{(\mathrm{UAV}), k}^j \cdot P_k) / (W \cdot \sigma^2))$ is the achievable data rate, while $T_{(\mathrm{UAV}), k}^j \in [0, \mu]$ denotes the transmission state at a decision epoch $j$.
%
%and evolves as $T_{(\mathrm{UAV}), k}^{j + 1} = T_{(\mathrm{UAV}), k}^j - R_{(\mathrm{UAV}), k}^j \cdot \min\{T_{(\mathrm{UAV}), k}^j / R_{(\mathrm{UAV}), k}^j, \delta\}$.
%
%After receiving all input data of a computation task, the UAV proceeds to process by creating a VM at the beginning of the subsequent decision epoch.
%
Let $\mathcal{K}^j \subseteq \mathcal{K}$ represent the subset of MUs, whose computation tasks are being simultaneously processed by the corresponding VMs at the UAV during a decision epoch $j$.
Denote by $C_0$ the computation service rate of a VM at the UAV given that the task is run in isolation, the degraded computation rate of each MU $k \in \mathcal{K}^j$ is modeled as $C^j = C_0 \cdot (1 + \varphi)^{1 - |\mathcal{K}^j|}$, where $|\cdot|$ means the cardinality of a set and $\varphi \in \mathds{R}_+$ is a factor specifying the percentage of reduction in the computation rate of a VM when multiplexed with another VM.
Accordingly, we obtain the remote processing delay of MU $k$ during decision epoch $j$ as $D_{(\mathrm{UAV}), k}^j = \min\{S_{(\mathrm{UAV}), k}^j / C^j, \delta\}$ with the remote processing state $S_{(\mathrm{UAV}), k}^j \in [0, \mu]$ showing the amount of input data to be processed at the beginning of an epoch $j$.
%
%the dynamics of which follows $I_{(\mathrm{MU}), k}^{j + 1} = I_{(\mathrm{MU}), k}^j - C^j \cdot D_{(\mathrm{UAV}), k}^j$.

\section{Problem Formulation and Game-Theoretic Solution}
\label{prob}

During each decision epoch $j$, the local state of a MU $k \in \mathcal{K}$ can be described by $\bm\xi_k^j = (L_{(\mathrm{MU}), k}^j, L_{(\mathrm{UAV})}^j, Q_k^j, I_k^j, S_{(\mathrm{MU}), k}^j,$ $S_{(\mathrm{UAV}), k}^j, T_{(\mathrm{BS}), k}^j, T_{(\mathrm{UAV}), k}^j) \in \mathcal{Z}$, where $\mathcal{Z}$ is a common finite state space for all MUs.
We use $\bm\xi^j = (\bm\xi_k^j, \bm\xi_{- k}^j) \in \mathcal{Z}^{|\mathcal{K}|}$ to represent the global system state with $- k$ denoting all the other MUs in $\mathcal{K}$ without the presence of a MU $k$.
Let $\pi_k$ be the stationary task scheduling policy employed by MU $k$.
When deploying $\pi_k$, MU $k$ observes $\bm\xi^j$ at the beginning of a decision epoch $j$ and accordingly, makes local as well as remote task scheduling decisions, that is, $\pi_k(\bm\xi^j) = (X_k^j, F_k^j)$.
We define an immediate utility function\footnote{To stabilize the training process of the proactive algorithm designed in this work, we choose an exponential function for the definition of an immediate utility, whose value does not dramatically diverge.}
\begin{align}\label{utilFunc}
  u_k\!\left(\bm\xi^j, \left(X_k^j, F_k^j\right)\right) = \exp\!\left(- D_k^j\right) + \eta \cdot \exp\!\left(- E_k^j\right),
\end{align}
to measure the satisfaction of experienced delay and consumed energy for each MU $k$ during each epoch $j$, where $\eta \in \mathds{R}_+$ is the weighting constant,
$D_k^j = D_{(\mathrm{MU}), k}^j + D_{(\mathrm{UAV}), k}^j + Y_{(\mathrm{BS}), k}^j + Y_{(\mathrm{UAV}), k}^j + \delta \cdot \max\{Q_k^j - X_k^j - \mathds{1}_{\{F_k^j > 0\}}, 0\}$ is composed of not only the processing and transmission delay but also the task queueing delay, while $E_k^j = E_{(\mathrm{MU}), k}^j + E_{(\mathrm{BS}), k}^j + E_{(\mathrm{UAV}), k}^j$ constitutes the total local energy consumption.

Along with the discussions, it can be easily verified that the randomness lying in a sequence of the global system states over the time horizon $\{\bm\xi^j: j \in \mathds{N}_+\}$ is Markovian.
Given a stationary task scheduling policy $\pi_k$ by each MU $k \in \mathcal{K}$ and an initial global state $\bm\xi^1 = \bm\xi \in \mathcal{Z}^{|\mathcal{K}|}$, we express the expected long-term discounted utility function $V_k(\bm\xi, (\pi_k, \bm\pi_{- k}))$ of MU $k$ as
\begin{align}\label{statValu}
  & V_k\!\left(\bm\xi, (\pi_k, \bm\pi_{- k})\right) =                                                                        \\
  & (1 - \gamma) \cdot \textsf{E}_{\left(\pi_k, \bm\pi_{- k}\right)}\!\!\left[\sum_{j = 1}^\infty (\gamma)^{j  - 1} \cdot
    u_k\!\left(\bm\xi^j, \left(X_k^j, F_k^j\right)\right) | \bm\xi^1 = \bm\xi\right]\!\!,                                    \nonumber
\end{align}
where $\gamma \in [0, 1)$ is the discount factor and the expectation $\textsf{E}_{(\pi_k, , \bm\pi_{- k})}[\cdot]$ is taken over different decision makings under different global system states following a joint task scheduling policy $(\pi_k, \bm\pi_{- k})$ across the decision epochs.
When $\gamma$ approaches $1$, (\ref{statValu}) approximates the expected long-term un-discounted utility as well \cite{Adel08}.
$V_k(\bm\xi, (\pi_k, \bm\pi_{- k}))$ is also termed as the state value function in a global system state $\bm\xi$ under a joint task scheduling policy $(\pi_k, , \bm\pi_{- k})$ \cite{Rich98}.

Due to the shared I/O resource at the UAV and the dynamic nature in networking environment, we formulate the problem of resource awareness among multiple MUs across the decision epochs as a non-cooperative stochastic game, in which the MUs are the players and there are a set $\mathcal{Z}^{|\mathcal{K}|}$ of global system states and a collection of task scheduling policies $\{\pi_k: \forall k \in \mathcal{K}\}$.
The aim of each MU $k$ is to device a best-response policy $\pi_k^*$ that maximizes $V_k(\bm\xi, (\pi_k, \bm\pi_{- k}))$, which can be formally formulated as $\pi_k^* = \arg\max_{\pi_k} V_k(\bm\xi, (\pi_k, \bm\pi_{- k}))$, $\forall \bm\xi \in \mathcal{Z}^{|\mathcal{K}|}$.
A NE, which is a best-response task scheduling policy profile $(\pi_k^*: k \in \mathcal{K})$, describes the rational behaviours of the MUs in a stochastic game \cite{Fink64}.
In order to operate the NE, a MU has to know the complete global system dynamics, which is prohibited in a non-cooperative networking environment \cite{Chen19}.
Define $\mathds{V}_k(\bm\xi) = V_k(\bm\xi, (\pi_k^*, \bm\pi_{-k}^*))$ as the optimal state-value function.

\section{Proactive DRL with Local Observations}
\label{probSolv}

In this section, we shall develop a proactive DRL algorithm to approach the NE task scheduling policy.

\subsection{Approximation from Local Observations}
\label{optiSolu}

During the competitive interactions with other MUs in the stochastic game, it is challenging for a MU to obtain the global system state information.
There still exists the possibility for each MU $k \in \mathcal{K}$ to acquire the side information, which is the partial observation $O_k^j$, of $\bm\xi_{- k}^j$ during a decision epoch $j$.
In this work, the partial observation of MU $k$ at the beginning of a decision epoch $j$ indicates the remote processing delay at the UAV from the previous epoch $j - 1$, namely, $O_k^j = D_{(\mathrm{UAV}), k}^{j - 1}$.
Therefore, $(\ref{statValu})$ can be approximated by (\ref{apprValu}),
\begin{figure*}[t]
\begin{align}\label{apprValu}
    V_k\!\left(\left(\bm\xi_k, O_k\right), (\pi_k, \bm\pi_{- k})\right) =
    (1 - \gamma) \cdot\textsf{E}_{\left(\pi_k, \bm\pi_{- k}\right)}\!\!\left[\sum_{j = 1}^\infty (\gamma)^{j  - 1} \cdot
    u_k\!\left(\bm\xi^j, \left(X_k^j, F_k^j\right)\right) | \left(\bm\xi_k^1, O_k^1\right) =
    \left(\bm\xi_k, O_k\right)\right]
\end{align}
\hrule
\end{figure*}
where $O_k^1$ is the initial partial observation of $\bm\xi_{- k}$.
Each MU $k$ then switches to solve the following single-agent MDP,
\begin{equation}\label{apprBestResp}
  \pi_k^* = \underset{\pi_k}{\arg\max}~ V_k\!\left(\left(\bm\xi_k, O_k\right), (\pi_k, \bm\pi_{- k})\right), \forall (\bm\xi_k, O_k).
\end{equation}
A dynamic programming approach to (\ref{apprBestResp}) based on the value or policy iteration requires complete a priori knowledge of the local state and observation transition statistics \cite{Rich98}.
The Q-learning enables each MU $k$ to learn $\pi_k^*$ in an unknown MEC system.
Define
\begin{align}\label{stat_acti_q1}
 & \mathds{Q}_k((\bm\xi_k, O_k), (X_k, F_k)) = (1 - \gamma) \cdot u_k(\bm\xi, (X_k, F_k)) +                                                 \nonumber\\
 & \gamma \cdot \sum_{\bm\xi_k, O_k} \mathbb{P}((\bm\xi_k', O_k') | (\bm\xi_k, O_k), (X_k, F_k)) \cdot \mathds{V}_k(\bm\xi_k', O_k'),
\end{align}
as the Q-function, where $X_k$ and $F_k$ are the decision makings at a current decision epoch, $\bm\xi_k'$ and $O_k'$ are the local state and the partial observation at the subsequent epoch, while $\mathds{V}_k(\bm\xi_k', O_k') = V_k((\bm\xi_k, O_k), (\pi_k^*, \bm\pi_{- k}^*))$.
In turn, $\mathds{V}_k(\bm\xi_k, O_k)$ can be straightforwardly obtained from
\begin{align}\label{stat_acti_q2}
    \mathds{V}_k(\bm\xi_k, O_k) = \max_{X_k, F_k} \mathds{Q}_k((\bm\xi_k, O_k), (X_k, F_k)).
\end{align}
By substituting (\ref{stat_acti_q2}) back into (\ref{stat_acti_q1}), we get (\ref{stat_acti_q3}),
\begin{figure*}[t]
\begin{align}\label{stat_acti_q3}
      \mathds{Q}_k((\bm\xi_k, O_k), (X_k, F_k))
  & = (1 - \gamma) \cdot u_k(\bm\xi, (X_k, F_k))                                                            \nonumber\\
  & + \gamma \cdot \sum_{\bm\xi_k, O_k} \mathbb{P}((\bm\xi_k', O_k') | (\bm\xi_k, O_k), (X_k, F_k)) \cdot
      \max_{X_k', F_k'} \mathds{Q}_k((\bm\xi_k', O_k'), (X_k', F_k'))
\end{align}
\hrule
\end{figure*}
with $X_k'$ and $F_k'$ denoting the local and remote computation task scheduling decisions under $(\bm\xi_k', O_k')$.

During the process of Q-learning, each MU $k \in \mathcal{K}$ in the network first observes $(\bm\xi_k, O_k) = (\bm\xi_k^j, O_k^j)$, $(X_k, F_k) = (X_k^j, F_k^j)$, $u_k(\bm\xi, X_k, F_k)$ at a current decision epoch $j$ as well as $(\bm\xi_k', O_k') = (\bm\xi_k^{j + 1}, O_k^{j + 1})$ at the next epoch $j + 1$, and then updates the Q-function iteratively as in (\ref{QLearRule}),
\begin{figure*}[t]
\begin{align}\label{QLearRule}
     \mathds{Q}_k^{j + 1}((\bm\xi_k, O_k), (X_k, F_k))
 & = \left(1 - \alpha^j\right) \cdot \mathds{Q}_k^j((\bm\xi_k, O_k), (X_k, F_k))                                  \nonumber\\
 & + \alpha^j \cdot \left((1 - \gamma) \cdot u_k(\bm\xi, (X_k, F_k)) + \gamma \cdot \max_{X_k', F_k'}
     \mathds{Q}_k^{j + 1}((\bm\xi_k, O_k), (X_k', F_k'))\right)
\end{align}
\hrule
\end{figure*}
where $\alpha^j \in [0, 1)$ is the learning rate.
It has been well established that if: 1) the global system state transition probability under $(\pi_k^*, \bm\pi_{- k}^*)$ is time-invariant; 2) $\sum_{j = 1}^\infty \alpha^j$ is infinite and $\sum_{j = 1}^\infty (\alpha^j)^2$ is finite; and 3) all $((\bm\xi_k, O_k), (X_k, F_k))$-pairs are visited infinitely often, the learning process converges towards $\pi_k^*$ \cite{Rich98}.

\subsection{Proactive DRL for NE Control Policy}
\label{learPoli}

We can easily find that for the system model being investigated in this paper, the joint space of local states and partial observations faced by each MU is extremely huge.
The tabular nature in representing the Q-function values makes the Q-learning impractical.
Inspired by the widespread success of a deep neural network \cite{Mnih15}, we adopt a double deep Q-network (DQN) to model the Q-function of a MU \cite{Hass16}.
However, the accuracy of (\ref{apprBestResp}), which is based on partial observations of other MUs in the MEC system, can be, in general, arbitrarily bad.
In order to overcome such a challenge from partial observability, we propose a slight modification to the DQN architecture.
That is, we replace the first fully-connected layer of the DQN with a long short-term memory (LSTM) layer \cite{Hoch97}, resulting in a deep recurrent Q-network (DRQN) \cite{Hauk15, Chen19A}.

More specifically, for each MU $k \in \mathcal{K}$ in the MEC system, $\mathds{Q}_k((\bm\xi_k, O_k), (X_k, F_k))$ is replaced by $\mathds{Q}_k(\mathbf{n}_k, (X_k, F_k); \bm\theta_k)$, where $\bm\theta_k$ contains a vector of parameters associated with the DRQN while $\mathbf{n}_k = \mathbf{n}_k^j$ consists of the $N$ most recent local states and partial observations up to a current decision epoch $j$, namely,
\begin{align}
  \mathbf{n}_k^j = \left(\left(\bm\xi_k^{j - n + 1}, O_k^{j - n + 1}\right): n = N, N - 1, \cdots, 1\right).
\end{align}
It is worth mentioning that $\mathbf{n}_k$ is taken as an input to the LSTM layer of the DRQN of MU $k$ for a proactive and more precise prediction of the current global system state $\bm\xi$.
Eventually, a MU leans the parameters of a DRQN, instead of finding the Q-function according to the rule in (\ref{QLearRule}).

\subsection{Offline Training by Digital Twin}
\label{offlTrai}

Simply being equipped with an independent DRQN at each MU raises two new technical challenges:
\begin{enumerate}
  \item the possibly asynchronous training of DRQNs at the MUs constrains the overall system performance; and
  \item in practice, the limited computation capability at the mobile device of a MU hinders the feasibility of training a DRQN locally.
\end{enumerate}
As a promising alternative, we set up a digital twin of the MEC system to offline train the DRQNs, the parameters of which can be preloaded to a MU during the network initiation.
From the assumptions made in this paper and the definition of an identical utility function structure as in (\ref{utilFunc}), the homogeneous behaviours in all MUs provide an opportunity for the digital twin to train a common DRQN with parameters $\bm\theta$.
In other words, we derive for each MU $k \in \mathcal{K}$, $\mathds{Q}_k(\mathbf{n}_k, (X_k, F_k); \bm\theta_k) = \mathds{Q}(\mathbf{n}_k, (X_k, F_k); \bm\theta)$.

To implement the DRQN offline training at the digital twin, we maintain a replay memory $\mathcal{M}^j$ to store the most recent $M$ experiences $\{\mathbf{m}^{j - M + 1}, \cdots, \mathbf{m}^j\}$ up to the beginning of each decision epoch $j$, where an experience $\mathbf{m}^{j - m + 1}$ ($m \in \{1, \cdots, M\}$) is given as (\ref{expe}).
\begin{figure*}[t]
\begin{align}\label{expe}
   \mathbf{m}^{j - m + 1} =
   \left(\left(\left(\bm\xi_k^{j - m}, O_k^{j - m}\right), \left(X_k^{j - m}, F_k^{j - m}\right),
              u_k\!\left(\bm\xi^{j - m}, \left(X_k^{j - m}, F_k^{j - m}\right)\right),
              \left(\bm\xi_k^{j - m + 1}, O_k^{j - m + 1}\right)\right): k \in \mathcal{K}\right)
\end{align}
\hrule
\end{figure*}
Meanwhile, a pool $\mathcal{N}^j = \{\mathbf{n}_k^j: k \in \mathcal{K}\}$ of $N$ latest local states and partial observations is kept to predict the global system state $\bm\xi^j$ for task scheduling policy evaluation at epoch $j$.
Both $\mathcal{M}^j$ and $\mathcal{N}^j$ are refreshed over the decision epochs.
We first randomly sample a mini-batch $\tilde{\mathcal{M}}^j = \{\breve{\mathcal{M}}^{j_1}, \cdots, \breve{\mathcal{M}}^{j_{\tilde{M}}}\}$ of size $\tilde{M}$ from $\mathcal{M}^j$, where each $\breve{\mathcal{M}}^{j_{\tilde{m}}} \nsubseteq \mathcal{M}^j$ ($\tilde{m} \in \{1, \cdots, \tilde{M}\}$) is given by (\ref{batcSamp}).
\begin{figure*}[t]
\begin{align}\label{batcSamp}
   \breve{\mathcal{M}}^{j_{\tilde{m}}} =
   \left\{\left(\mathbf{n}_k^{j_{\tilde{m}}}, \left(X_k^{j_{\tilde{m}}}, F_k^{j_{\tilde{m}}}\right),
                u_k\!\left(\bm\xi^{j_{\tilde{m}}}, \left(X_k^{j_{\tilde{m}}}, F_k^{j_{\tilde{m}}}\right)\right),
                \mathbf{n}_k^{j_{\tilde{m}} + 1}\right): k \in \mathcal{K}\right\}
\end{align}
\hrule
\end{figure*}
Then the set $\bm\theta^j$ of parameters at epoch $j$ is updated by minimizing the accumulative loss function, which is defined as in (\ref{lossFunc}),
\begin{figure*}[t]
\begin{align}\label{lossFunc}
 & \textsf{LOSS}\!\left(\bm\theta^j\right) =
   \textsf{E}_{\left\{\breve{\mathcal{M}}^{j_{\tilde{m}}} \in \tilde{\mathcal{M}}^j\right\}}\!
   \left[\left(\sum_{k \in \mathcal{K}} \left((1 - \gamma) \cdot
   u_k\!\left(\bm\xi^{j_{\tilde{m}}}, \left(X_k^{j_{\tilde{m}}}, F_k^{j_{\tilde{m}}}\right)\right) +
   \vphantom{\underset{X_k^{j_{\tilde{m}} + 1}, F_k^{j_{\tilde{m}} + 1}}{\arg\max}}\right.\right.
   \vphantom{\left(\underset{X_k^{j_{\tilde{m}} + 1}, F_k^{j_{\tilde{m}} + 1}}{\arg\max}\right)^2}\right.                                   \nonumber\\
 & \left.\left.\left.\gamma \cdot \mathds{Q}\Bigg(\mathbf{n}_k^{j_{\tilde{m}} + 1},
                                  \underset{X_k^{j_{\tilde{m}} + 1}, F_k^{j_{\tilde{m}} + 1}}{\arg\max}
                                  \mathds{Q}\!\left(\mathbf{n}_k^{j_{\tilde{m}} + 1}, \left(X_k^{j_{\tilde{m}} + 1}, F_k^{j_{\tilde{m}} + 1}\right); \bm\theta^j\right); \bm\theta_{-}^j\Bigg) -
   \mathds{Q}\!\left(\mathbf{n}_k^{j_{\tilde{m}}}, \left(X_k^{j_{\tilde{m}}}, F_k^{j_{\tilde{m}}}\right); \bm\theta^j\right)\right)\right)^2\right]
\end{align}
\hrule
\end{figure*}
where $\bm\theta_{-}^j$ is the set of parameters of the target DRQN at a certain previous decision epoch before epoch $j$.

\section{Numerical Experiments}
\label{simu}

In order to quantify the performance gain from the proposed proactive DRL scheme in a UAV-assisted MEC system, numerical experiments based on TensorFlow \cite{Abad16} are conducted.
For experimental purpose, we build up a terrestrial MEC system, which is with $B = 4$ BSs in a $0.4\times0.4$ Km$^2$ square area.
The BSs are placed at equal distance apart, and the square area is divided into $|\mathcal{L}| = 1600$ locations with each representing a small area of $10\times10$ m$^2$.
The channel model in \cite{Chen19} and the LOS model in \cite{Zeng16} are assumed, respectively, for $G_{b, k}^j$ and $G_{(\mathrm{UAV}), k}^j$, $\forall k \in \mathcal{K}$, $\forall b \in \mathcal{B}$ and $\forall j$.
We use the mobility configurations as in \cite{Xi19} for the MUs and the UAV.
Regarding the DRQN, we design two fully connected layers after the LSTM layer with each of the three layers containing $32$ neurons.
ReLU is selected as the activation function \cite{Nair10} and Adam as the optimizer \cite{King15}.
Other parameter values are listed in Table \ref{tabl1}.
\begin{table}[t]
  \caption{Parameter values in experiments.}\label{tabl1}
        \begin{center}
        \begin{tabular}{|c|c||c|c|}
              \hline
              Parameter       & Value                 & Parameter     & Value                             \\\hline
              \hline
              $\mu$           & $500$ Kbits           & $\vartheta$   & $1300$                            \\\hline
              $H$             & $100$ meters          & $W$           & $1$ MHz                           \\\hline
              $\sigma^2$      & $-174$ dBm/Hz         & $\delta$      & $10^{-2}$ second                  \\\hline
              $P_k$           & $3$ Watt, $\forall k$ & $\eta$        & $3$                               \\\hline
              $\rho$          & $2$ GHz               & $\varphi$     & $0.1$                             \\\hline
              $\zeta$         & $10^{-3}$ second      & $C_0$         & $2\cdot 10^7$ bits/second         \\\hline
              $\tau$          & $2.5 \cdot 10^{-28}$  & $N$           & $50$                              \\\hline
              $M$             & $5000$                & $\breve{M}$   & $200$                             \\
              \hline
        \end{tabular}
        \end{center}
\end{table}

For the performance comparisons, we design the following four baseline schemes as well.
\begin{enumerate}
  \item \emph{Local Computation} -- Each MU processes locally all arriving computation tasks.
  \item \emph{Cloud Execution} -- All arriving computation tasks at the MUs are offloaded to the MEC cloud for execution via the BS with the best channel gain.
  \item \emph{UAV Execution} -- All queued computation tasks from the MUs are processed by the VMs at the UAV.
  \item \emph{Greedy Processing} -- Each MU schedules the local task computation or offloads the computation to the UAV or the MEC cloud whenever possible.
\end{enumerate}

\begin{figure}[t]
  \centering
  \includegraphics[width=21pc]{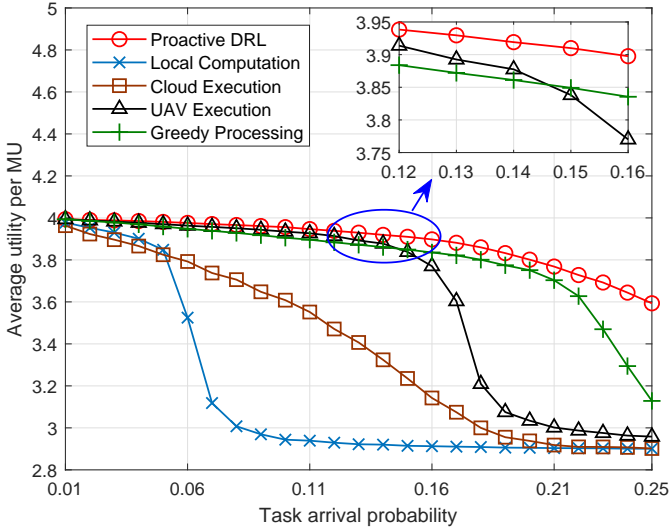}
  \caption{Average utility performance per MU across the decision epochs versus computation task arrival probability $\lambda$.}
  \label{simu01}
\end{figure}

In the experiments, the priority is to demonstrate the average utility performance per MU across the decision epochs from the proposed proactive DRL scheme and the four baselines under various computation task arrival probabilities.
We assume $|\mathcal{K}| = 12$ MUs in the MEC system.
The results are depicted in Fig. \ref{simu01}.
It can be observed from the curves that the average utility performance from the proposed scheme, the local computation, the cloud execution, the UAV execution and the greedy processing deceases as the computation task arrival probability $\lambda$ increases, which is in accordance with our intuition lying in the surge of per-MU task queue length.
Due to the LOS wireless transmissions between the MUs and the UAV, the UAV execution scheme achieves better average utility performance than the cloud execution scheme.
As $\lambda$ increases, more task computations are offloaded for UAV execution under the greedy processing scheme to avoid the possible handover delay, though the cloud execution scheme outperforms the local computation scheme.
Among the four baselines, the greedy processing scheme exhibits the best performance under large values of $\lambda$.
Last but not least, the results clearly show that the proposed scheme provides a significant performance gain, compared with the four baselines.

\section{Conclusions}
\label{conc}

In this work, our focus is to study the design of a stochastic local and remote computation scheduling policy for each MU in a UAV-assisted MEC system, which takes into account the system dynamics originated from the UAV and the MU mobilities as well as the time-varying computation task arrivals.
The non-cooperative interactions among the MUs across the decision epochs are formulated as a stochastic game.
To approach the NE, we derive a proactive DRL scheme, with which each MU schedules local and remote computations using only the local information.
The homogeneity in the behaviours of MUs facilitates the use of a digital twin to offline train the proposed scheme.
From numerical experiments, we find that compared with the four baselines, the proposed proactive DRL scheme achieves the best average utility performance.


\end{document}